%% file: main.tex
\documentclass[journal]{IEEEtran}
\usepackage{balance}

\title{{WebAPIRec}: Recommending {Web} APIs to Software Projects via Personalized Ranking}

\usepackage{color}

\ifCLASSINFOpdf
\else
\fi
\usepackage[cmex10]{amsmath}
\usepackage{algorithm}
\usepackage{algorithmic}
\usepackage{booktabs}
\usepackage{url}
\usepackage{graphicx}
\usepackage{multirow}

\begin{document}



%
\author{
\IEEEauthorblockN{Ferdian Thung, Richard J. Oentaryo, David Lo, and Yuan Tian}
\IEEEauthorblockA{\\School of Information Systems, Singapore Management University, Singapore\\
Email: \{ferdiant.2013, roentaryo, davidlo, yuan.tian.2012\}@smu.edu.sg}}


\maketitle

\begin{abstract}
Application programming interfaces (APIs) offer a plethora of functionalities for developers to reuse without reinventing the wheel. Identifying the appropriate APIs given a project requirement is critical for the success of a project, as many functionalities can be reused to achieve faster development. However, the massive number of APIs would often hinder the developers' ability to quickly find the right APIs. In this light, we propose a new, automated approach called {\tt {WebAPIRec}} that takes as input a project profile and outputs a ranked list of {web} APIs that can be used to implement the project. At its heart, {\tt {WebAPIRec}} employs a personalized ranking model that ranks {web} APIs specific (personalized) to a project. Based on the historical data of {web} API usages, {\tt {WebAPIRec}} learns a model that minimizes the incorrect ordering of {web} APIs, i.e., when a used {web} API is ranked lower than an unused (or a not-yet-used) {web} API. We have evaluated our approach on a dataset comprising 9,883 {web} APIs and 4,315 web application projects from ProgrammableWeb with promising results. For 84.0\% of the projects, {\tt {WebAPIRec}} is able to successfully return correct APIs that are used to implement the projects in the top-5 positions. This is substantially better than the recommendations provided by ProgrammableWeb's native search functionality. {\tt {WebAPIRec}} also outperforms McMillan  \emph{et al.}'s application search engine and popularity-based recommendation. 
\end{abstract}



%
\IEEEpeerreviewmaketitle

\input{introduction}
\input{prelim}

\input{framework}
\input{method}
\input{feature}

\input{experiment}
\input{related}
\input{conclusion}
\section*{Acknowledgment}

This research was supported by the Singapore Ministry of Education (MOE) Academic Research Fund (AcRF) Tier 1 grant.


\bibliographystyle{IEEEtranS}
\bibliography{main}



%

\begin{IEEEbiography}[{\includegraphics[width=1in,height=1.25in,clip,keepaspectratio]{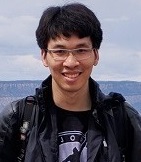}}]{Ferdian Thung}
	is a PhD student in the School of Information System,
	Singapore Management University. He started his PhD program in 2013. Previously,
	he received his bachelor degree in the School of Electrical Engineering and Informatics
	from Bandung Institute of Technology, Indonesia in 2011. He has worked as a research
	engineer for more than a year before joining the PhD program. His research interest is in software engineering and data mining area. He has
	been working on automated predictions, recommendations, and empirical studies in
	software engineering.
\end{IEEEbiography}
\begin{IEEEbiography}[{\includegraphics[width=1in,height=1.25in,clip,keepaspectratio]{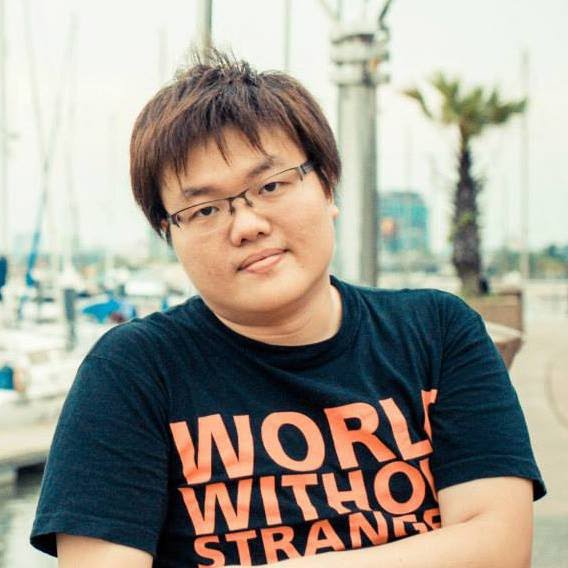}}]{Richard J. Oentaryo}
	is a Senior Data Scientist at McLaren Applied Technologies Singapore, a new R\&D arm of the McLaren Technology Group in Asia-Pacific region. Previously, he was a Research Scientist at the Living Analytics Research Centre (LARC), School of Information Systems, Singapore Management University (SMU) in 2011-2016, and a Research Fellow at the School of Electrical and Electronic Engineering, Nanyang Technological University (NTU) in 2010-2011. He received Ph.D. and B.Eng. (First Class Honour) degrees from the School of Computer Engineering (SCE), NTU, in 2011 and 2004 respectively. Dr. Oentaryo is a member of the Association for Computing Machinery (ACM) and Institute of Electrical and Electronics Engineers (IEEE). He has published in numerous international journals and conferences, and received such awards as the IES Prestigious Engineering Achievement Award in 2011, IEEE-CIS Outstanding Student Paper Travel Grant in 2006 and 2009, and ITMA Gold Medal cum Book Prize in 2004.
\end{IEEEbiography}
\begin{IEEEbiography}[{\includegraphics[width=1in,height=1.25in,clip,keepaspectratio]{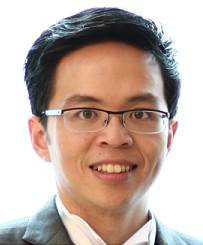}}]{David Lo}
	received his PhD degree from the School of Computing, National University of Singapore in 2008. He is currently an Associate Professor in the School of Information Systems, Singapore Management University. He has close to 10 years of experience in software engineering and data mining research and has more than 200 publications in these areas. He received the Lee Foundation Fellow for Research Excellence from the Singapore Management University in 2009, and a number of international research awards including several ACM distinguished paper awards for his work on software analytics. He has served as general and program co-chair of several established international conferences (e.g., IEEE/ACM International Conference on Automated Software Engineering), and editorial board member of a number of high-quality journals (e.g., Empirical Software Engineering).
\end{IEEEbiography}
\begin{IEEEbiography}[{\includegraphics[width=1in,height=1.25in,clip,keepaspectratio]{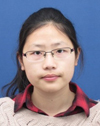}}]{Yuan Tian}
	is currently a PhD student in the School of Information Systems, Singapore Management University. She started her PhD program in 2012. Previously, she received her bachelor degree in the College
	of Computer Science and Technology from Zhejiang University, China in 2012. Her research is in software system and data mining area. Particularly, she is interested in analyzing textual information in software
	repositories.
\end{IEEEbiography}

\balance
\end{document}

%% file: introduction.tex
\section{Introduction}
\label{sec:introduction}

Developing a software project is not an easy task, as customers usually demand many features to be implemented. To aid their jobs, developers often use third party libraries that provide relevant functionalities through application programming interfaces (APIs)~\cite{raemaekers2012analysis}. APIs provide functionalities for certain tasks that can be (re)used by developers to expedite project developments. Using APIs prevents developers from reinventing the wheel, thus allowing them to focus on more important tasks at hand. Hence, it is usually a good idea to find suitable APIs and use them in a project. 
Moreover, by building upon existing APIs, features can be completed faster as many APIs are well designed and their functionalities have been tested by many client applications.

Finding the right APIs, however, is not as straightforward as it may seem. Thousands of APIs have been developed to cater for various purposes, and developers are often unaware of the existence of APIs suitable for a particular feature of the project that they are developing. Of course, some APIs are well known, but the majority of APIs do not enjoy such luxury~\cite{yu2009innovation}. {Moreover, although some API choices are obvious (e.g., if we want to add Facebook support, we do not have much choice except using Facebook API), the number of such obvious API choices is not many. In general, finding APIs for various needs, e.g., music management, typically involves many possible alternatives and the choice will largely depend on the project requirement. Some examples of music management APIs are MusicBrainz, Soundiiz, and Toma.hk. MusicBrainz can be used to extract music metadata, Soundiiz can be used to create music playlist, and Toma.hk can be used to play music from different sources. The choice of which API to use would depend on the need and requirement of a target application.} These facts necessitate the development of an automated recommendation system that can help developers find APIs that they need for their projects.


In this paper, we propose a new approach dubbed {\tt {WebAPIRec}} to recommend {web} APIs based on project profiles. In {\tt {WebAPIRec}}, we define a project profile as the textual description and keywords of the project. It is worth noting that our approach does not require the {web} API source code to be available. This requirement is important as many proprietary yet useful {web} APIs do not come with source code.
Examples of web APIs include Google Maps, Bing Maps, YouTube, and Last.fm, which are often used as key components in many projects. These {web} APIs offer essential functionalities and usually come with data that can be used to complete various features in a more efficient way. 

Given a new project profile, our approach recommends {web} APIs by analyzing past projects and the {web} APIs that they use. {\tt {WebAPIRec}} consists of two phases: {\em training} and {\em deployment phase}. In the training phase, {\tt {WebAPIRec}} analyzes past projects and their used {web} APIs to build a personalized ranking model that aims to minimize ranking errors in the training data. Personalized ranking means that the ranking of {web} APIs is specific to each project, and thus different projects have different {web} API rankings. A ranking error occurs if a {web} API used by some project is ranked lower than an unused {web} API. In the deployment phase, {\tt {WebAPIRec}} analyzes the profile of a new project using the trained model. It then assigns a relevancy score to each {web} API. A higher relevancy score implies that the API is deemed more relevant. Finally, {\tt {WebAPIRec}} ranks the {web} APIs in a descending order of their relevancy and returns a list of recommended {web} APIs. {This list is intended to help developers to pick web APIs more efficiently. It does not explicitly return a composition of web APIs for the project.



To illustrate the usefulness of our approach, consider the following scenario. A developer has no idea what {web} API to use for developing his application. Normally, he will surf the web to find a suitable {web} API. However, not all web pages are related to {web} APIs and, even if they are, he still needs to read the {web} API descriptions and decide whether each of them is usable or not. If he thinks a web API is usable, he will try the {web} API. Still, after trying it, the {web} API may not meet his expectations. There may be numerous trials and errors before he finds the {web} API that best matches his needs. We thus develop {\tt {WebAPIRec}} to provide an automated recommender system that can help reduce the effort needed by a developer to find the right {web} API. 



To validate our {\tt {WebAPIRec}} approach, we use the web application projects and web APIs extracted from the ProgrammableWeb website\footnote{http://www.programmableweb.com/}. This dataset has a total of 9,883 {web} APIs and 4,315 projects. 
We evaluate the effectiveness of our approach in terms of Hit@N, MAP@N, MAP, and MRR, which are popular metrics for evaluating recommender systems~\cite{shaowei/icpc2014,RaoK11,ZhouZL12,SahaLKP13,SunLWJK10,RAN07,ThungWLL13}. Our experiment shows that our approach achieves Hit@5, Hit@10, MAP@5, MAP@10, MAP, and MRR scores of 0.840, 0.880, 0.697, 0.687, 0.626, and 0.750, respectively. The Hit@5 score implies that for 84.0\% of the projects, {\tt {WebAPIRec}} can successfully return correct {web} APIs, which are used to implement the projects at the top-5 positions.

We have compared the effectiveness of our approach against the native search functionality of ProgrammableWeb. We input the profile of a project (in full or in part) and evaluate the list of libraries that the search functionality returns. However, we find that the search functionality is limited and it achieves only Hit@5, Hit@10, MAP@5, MAP@10, MAP, and MRR scores of at most 0.046, 0.047, 0.041, 0.042, 0.042, and 0.038 respectively. We have also compared our approach against several other baselines based on McMillan \emph{et al.}'s application search engine~\cite{mcmillan2012exemplar} and popularity-based recommendation.
We find that our approach outperforms all of them. The best performing baseline achieves significantly lower Hit@5, Hit@10, MAP@5, MAP@10, MAP, and MRR scores of 0.591, 0.675, 0.414, 0.417, 0.363, and 0.476 respectively. 
Comparing the Hit@5 scores of {\tt {WebAPIRec}} with those of the baselines, {\tt {WebAPIRec}} outperforms the best performing baseline by a substantial margin of 42.1\%.


We summarize our main contributions as follows:
\begin{enumerate}
 \item We propose a new approach named {\tt {WebAPIRec}} that recommends {web} APIs by analyzing past similar projects and {web} APIs that they use, and model the recommendation task as a ranking problem. To our best knowledge, {\tt {WebAPIRec}} is the first approach that employs a personalized ranking model to learn the correct ordering of {web} APIs for a specific project. Our approach recommends top-$k$  {web} APIs that can most likely be used to implement the project.

\item We have comprehensively evaluated our approach on a dataset extracted from ProgrammableWeb. Our experiment shows that {\tt {WebAPIRec}} is able to achieve satisfactory Hit@$N$, MAP, MAP@$N$ and MRR scores. These results are substantially better than the results for the ProgrammableWeb's native search functionality, McMillan \emph{et al.}'s application search, and popularity-based recommendation.
\end{enumerate}



%% file: prelim.tex
\section{Preliminaries}
\label{sec:prelim}


\subsection{ProgrammableWeb Dataset}\label{sec:programmableweb}


ProgrammableWeb is a website that collects information about APIs released as web services and web application projects that use them. It contains a collection of thousands of APIs implementing various functionalities. Table~\ref{tab:apiEx} shows the profile of an API in our dataset. The profile of an API contains several pieces of information such as its name, short description (i.e., summary), long description, and keywords (i.e., tags). In this paper, we refer to a merged text that contains the name, short description, and long description of an API as the {\em textual description} of the API. We represent each API by its textual descriptions and keywords. 


\begin{table}[!htb]
  \vspace{-0.4cm}
  \scriptsize
  \centering
  \caption{A Sample API Profile}\label{tab:apiEx}
  \vspace{-0.2cm}
  \begin{tabular}{lp{1.9in}}
  \toprule
  \multicolumn{2}{c}{\bf Last.fm API} \\
  \midrule
  {\bf Short Description} & Online audio service \\
  {\bf Long Description} & The Last.fm API gives users the ability to build programs using Last.fm data, whether on the web, the desktop or mobile devices. The RESTful API allows for read and write access to the full slate of last.fm music data resources - albums, artists, playlists, events, users, and more. It allows users to call methods that respond in either XML or JSON. \\
  {\bf Keywords} & music \\
  \bottomrule
  \end{tabular}
  \vspace{-0.2cm}
\end{table}

ProgrammableWeb contains thousands of web application projects. Table~\ref{tab:mashupEx} shows the profile of a project in our dataset. The profile contains several pieces of information including: a long description of the project and the relevant keywords (i.e., tags). A web application project does not have a short description in ProgrammableWeb. We refer to the long description of a web application project as its {\em textual description}. Similar to an API, we represent each web application project by its textual descriptions and keywords.





\begin{table}[!htb]
  \centering
  \vspace{-0.4cm}
  \scriptsize
  \caption{A Sample Web Application Project Profile}\label{tab:mashupEx}
  \vspace{-0.2cm}
  \begin{tabular}{lp{1.9in}}
  \toprule
  \multicolumn{2}{c}{\bf Ivy FM - Discover new music every day} \\
  \midrule
   {\bf Long Description} &  Discover new music every day with Ivy FM. It plays great songs continuously in each genre from the best artists in the world. Select your channel, listen great music, share it and enjoy.\\
   {\bf Keywords} & music, streaming\\
   {\bf APIs} & Last.fm, Youtube\\
  \bottomrule
  \end{tabular}
  \vspace{-0.4cm}
\end{table}

\subsection{IR \& NLP Techniques}\label{sec.ir}

{\tt {WebAPIRec}} make use of information retrieval (IR) and natural language processing (NLP) techniques. They include parts-of-speech (POS) tagging technique from NLP and text preprocessing, vector space model (VSM), and cosine similarity techniques from IR. We describe each of them below.

\subsubsection{Parts-of-Speech Tagging}

POS tagging is a natural language processing technique that assigns a part of speech label to every word in a textual document (in our case: a textual description of an API or a project). Common parts of speech include: noun, verb, adjective, adverb, etc. Various algorithms have been proposed to perform POS tagging. One of the most advanced family of POS tagging algorithms is stochastic POS taggers, which consider the context of a word to decide its POS tag~\cite{TreeTagger95,TnT2000,Stanford03}. In this work, we use the popular Stanford (stochastic) POS tagger~\cite{Stanford03}, which has also been used in many software engineering studies, e.g.,~\cite{Binkley11}.

\subsubsection{Text Preprocessing}

In this phase, we break a text data into a more suitable representation that can later be converted into an IR model. Also, since text data are often noisy (i.e., it contains many unimportant words, closely related words that are in different tenses, etc.), additional preprocessing steps are needed. In this work, the preprocessing steps are:

\begin{itemize}
 \item {\bf Tokenization.} It is a process of breaking a text document into its constituent word tokens. Delimiters, such as punctuation marks and white spaces, are used as boundaries between one word token and another. At the end of this process, each text document is represented by a bag (or multi-set) of word tokens.
 \item {\bf Stop Word Removal.} This involves removing words that appear very frequently and thus help very little in discriminating one document from another. Examples of these stop words include: ``I'', ``you'', ``are'', etc. In this work, we use the list of English stop words from \url{http://jmlr.org/papers/volume5/lewis04a/a11-smart-stop-list/english.stop}.
 \item {\bf Stemming.} It is a process of converting a word to its base form, typically by removing a suffix from the word. 
 For example, using stemming, words ``reads'' and ``reading'' would all be converted to ``read''. Without stemming, these words will be considered as different words altogether. We use the Porter stemming method~\cite{P80} to reduce each word to its stemmed form. 
\end{itemize}

\subsubsection{Vector Space Model}\label{sec.prelimVSM}

Text preprocessing will convert a textual \emph{document}---i.e., a project or API description---into a bag of words. In the bag of words representation, important words are not distinguished from unimportant ones. To consider the relative importance of words, IR researchers proposed the vector space model (VSM), which represents a textual document as a vector of weights~\cite{manning2008introduction}. Each weight corresponds to a word and indicates the relative importance of that word. VSM is constructed by analyzing many bags of words representing a set of documents in a \emph{corpus} (i.e., a collection of project or API descriptions).

Many weighting schemes can be used to infer the importance of a word. In this work, we use the popular term frequency-inverse document frequency (tf-idf) scheme~\cite{rajaraman2012mining}. This scheme is based on two intuitions. Firstly, words (terms) that appear frequently in a document are more important than words that appear rarely in it. For example, a document that has many occurrences of the word ``Texas'' is likely to be related to ``Texas''. Secondly, words that appear in many documents are less able to distinguish one document from another, and should be given a smaller weight. For example, if all documents in a corpus contains the word ``Software'', then this word is unimportant, as it cannot distinguish one document from another.

Given a document $D$ in a corpus $C$, we can compute the weight of every word that appears in $D$. To compute the term frequency (tf) of a word $d$ in a document $D$, we simply count how many times the word appear in $D$. To compute the inverse document frequency (idf) of a word $d$ in corpus $C$, we first compute the document frequency (df) of $d$, which is the number of documents in $C$ that contains $d$.  We then normalize this number by dividing it by the number of documents in $C$. The idf is simply the logarithm of the reciprocal of this normalized number. In turn, the tf-idf weight of a word $d$ is the product of its term frequency and inverse document frequency. Formally, the tf-idf weight of a word $d$ in a document $D$ of a corpus $C$ (denoted as $w_{(d,D,C)}$) is:
\begin{align}
w_{(d,D,C)} &= \mathit{TF}_{(d,D)} \times \mathit{IDF}_{(d,C)} \nonumber\\
            &= \mathit{TF}_{(d,D)} \times \log \left( \frac{N_C}{\mathit{DF}_{(d,C)}} \right)
\end{align}
where $\mathit{TF}_{(d,D)}$ refers to the term frequency of word $d$, $N_C$ refers to the number of documents in corpus C, and $\mathit{DF}_{(d,C)}$ refers to the document frequency of word $d$.

We denote the VSM representation of a document $D$ considering a corpus $C$ as $VSM_C(D)$. In our implementation, we use a sparse matrix representation for the API and project documents (i.e., we only store the non-zero entries).

\subsubsection{Cosine Similarity}\label{sec.cosine}

To compute the similarity of two documents, we can take their VSM representations and compare the two vectors of weights by computing their cosine similarity~\cite{manning2008introduction}. Consider two vectors $a$ and $b$ of size $N$; their cosine similarity is:
\begin{align}
    \mathit{Sim}(a,b) = \frac{\sum\nolimits_{i=1}^N w_{i,a} \times w_{i,b}}{\sqrt{\sum\nolimits_{i=1}^N w_{i,a}^2}\sqrt{\sum\nolimits_{i=1}^N w_{i,b}^2}}
\end{align}
where $w_{i,a}$ refers to the $i^{th}$ weight in vector $a$.

%
%
%


%% file: framework.tex
\section{API Recommendation System}
\label{sec:WebAPIRec}



\begin{figure}[!t]
\centering
\hspace{-0.5cm}\includegraphics[width=3.5in]{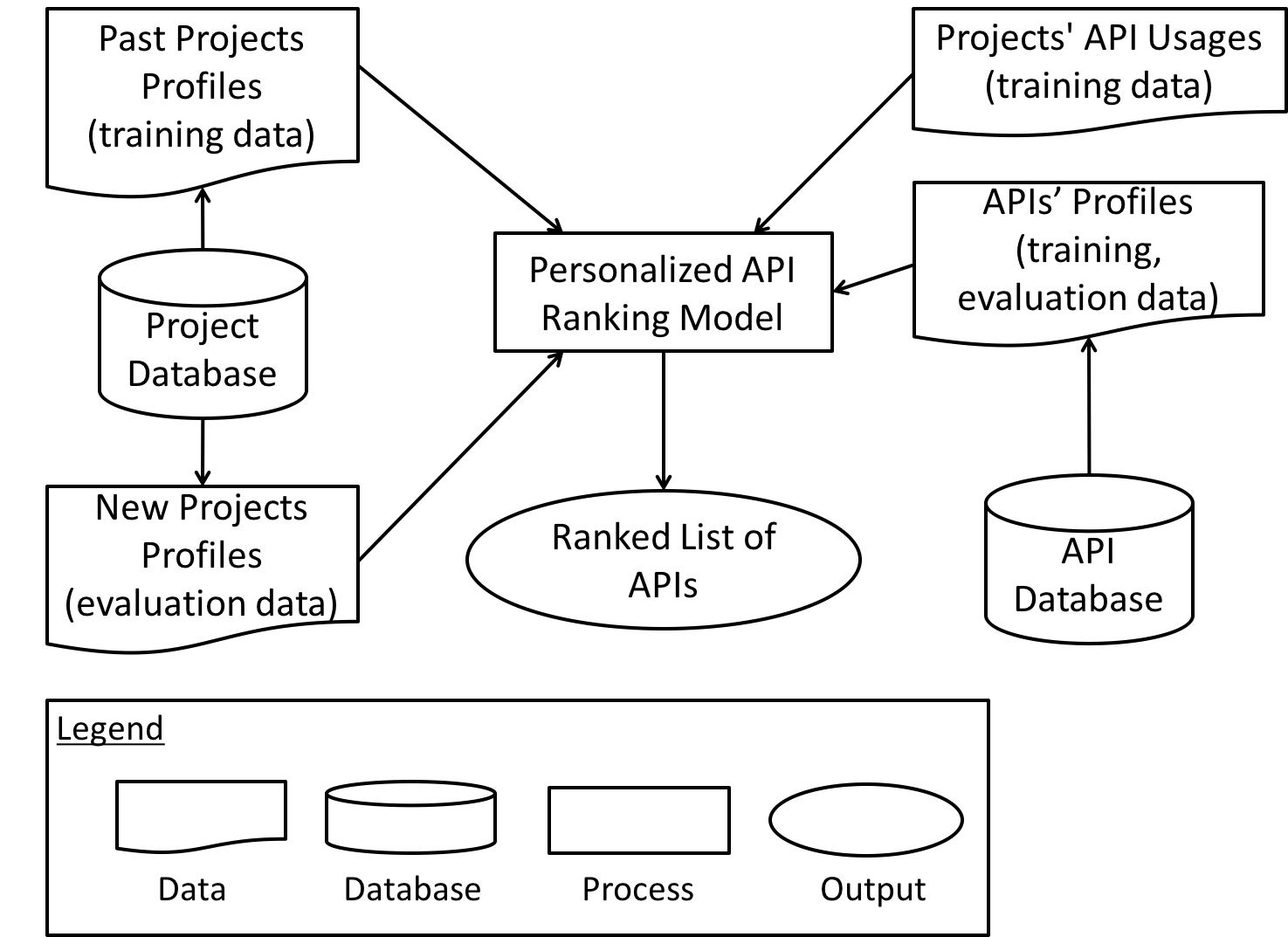}
\vspace{-0.3cm}
\caption{Architecture of {\tt {WebAPIRec}}}
\label{fig:framework}
\vspace{-0.5cm}
\end{figure}

The architecture of {\tt {WebAPIRec}} is outlined in Figure~\ref{fig:framework}. It takes as input: a new project profile, a set of API profiles, and a set of past projects. From the new project profile and each API profile, {\tt {WebAPIRec}} takes its textual descriptions and keywords. From each past project, {\tt {WebAPIRec}} takes its textual descriptions, keywords, and APIs that was used. {\tt {WebAPIRec}} analyzes these inputs and finally produces a ranked list of APIs to be recommended to the target project.

{\tt {WebAPIRec}} has two operating phases: {\em training phase} and {\em deployment phase}. In the former phase, {\tt {WebAPIRec}} takes as input a set of API profiles and a set of past projects along with the APIs that they use. It then learns a personalized ranking model (see Section~\ref{sec:method}). In the deployment phase, it takes as input the new project profile, a set of API profiles, and the trained personalized API ranking model. It then applies the model to the new project profile and outputs a ranked list of recommended APIs.



To train the personalized ranking model in the {\em training phase}, {\tt {WebAPIRec}} needs to represent the profile of each past project (i.e., training data) as a feature vector. 
{\tt {WebAPIRec}} first identifies nouns from the textual descriptions using the Stanford POS tagger. These nouns carry more meaning than other kinds of words, as advocated in~\cite{Capobianco12,Shokripour13}. {\tt {WebAPIRec}} then combines the extracted nouns with the keywords, remove stop words, stem each of the remaining words, and construct a VSM feature vector. The same process can be done to convert an API profile into a feature vector. 
These project and API feature vectors are then used to construct a set of training triples $(p,a,a')$, which serves as input to the personalized ranking model. In a triple $(p,a,a')$, $p$ is the feature vector of a project in the training data, $a$ is the feature vector of an API that is used by project $p$, and $a'$ is the feature vector of an API \emph{not} used by project $p$. At the end of the training phase, the ranking model will have learned how to rank a list of APIs based on their feature vectors and the feature vector of the target project.

In the {\em deployment phase}, similar to the {\em training phase}, {\tt {WebAPIRec}} first constructs feature vectors from a new project profile and API profiles. Using the learned personalized API ranking model, {\tt {WebAPIRec}} computes the relevancy of each API and sort the APIs (in descending order) based on these scores. The sorted APIs are output as a list of recommended APIs.


%

%% file: method.tex
\section{Personalized Ranking}
\label{sec:method}

{\tt {WebAPIRec}} casts the API recommendation problem as a \emph{personalized ranking} task. Under this formulation, our goal is to provide a ranked list of APIs that are specific (i.e., personalized) to each project. Specifically, we consider the setting where {\tt {WebAPIRec}} takes as input a set of training triples ($p$,$a$,$a'$) where $p$ is a feature vector of a project, $a$ is a feature vector of an API library used in $p$, and $a'$ is a feature vector of an API not used in $p$. Based on these training triples, a personalized ranking model learns how to rank APIs for a target project by jointly utilizing their feature vectors.


\subsection{Notation and Desiderata}


We first define our notations here. Let $P$ be the set of all software projects and $A$ the set of all web APIs. Accordingly, the recommendation task is to provide a specific project $p \in P$ with a total ordering $>_p$ of all APIs $a \in A$. Essentially, a sound ranking $>_p$ requires several criteria to be fulfilled:
\begin{align}
\label{eqn:totality}
\forall a, a' \in A \colon & a \neq a' \Rightarrow a >_p a' \vee a' >_p a\\
\label{eqn:antisym}
\forall a, a' \in A \colon & a >_p a' \wedge a' >_p a \Rightarrow a = a'\\
\label{eqn:transitivity}
\forall a, a', a'' \in A \colon & a >_p a' \wedge a' >_p a'' \Rightarrow a >_p a''
\end{align}

\noindent The formulae (\ref{eqn:totality})--(\ref{eqn:transitivity}) correspond to the so-called \emph{totality} (i.e., $a$ and $a'$ should be comparable), \emph{anti-symmetry} (i.e., unless $a=a'$, $a$ and $a'$ should have different ranks), and \emph{transitivity} properties (i.e., if $a$ ranks higher than (or equal to) $a'$ and $a'$ ranks higher than (or equal to) $a''$, then $a$ ranks higher than (or equal to) $a''$), respectively~\cite{davey2002introduction}.


The personalized ranking model will in turn learn to rank APIs based on a set of training triples $D$:
\begin{align}
\label{eqn:pair_data}
D = \{(p,a,a') | a \in A_p \wedge a' \in A \backslash A_p \}
\end{align}
\noindent where $A_p$ refers to the set of APIs used by a project $p$, and each element/triple $(p,a,a') \in D$ implies that project $p$ prefers API $a$ over API $a'$. 

\subsection{Ranking Model}

Our personalized ranking model computes a compatibility score between a project $p$ and an API $a$. Specifically, for any $(p,a)$ pair, our model defines the compatibility score $f(p,a)$ as a weighted sum of $J$ interaction features:
\begin{align}
f(p,a) = &\sum_{j=1}^J \theta_j x_j(p,a)
\end{align}
where each feature $x_j(p,a)$ quantifies a specific type of interaction between the project $p$ and API $a$, and $\theta_j$ is the weight parameter to be identified by the training procedure. Further details on which features $x_j(p,a)$ we use in the recommendation task will be given later in Section \ref{sec:feature}.

After training is completed, we can compute for a new project $p'$ the score $f(p',a)$  using the identified weight parameters $\theta_j$ and feature $\theta_j(p',a)$. We may then sort the scores $f(p',a)$ computed for all APIs $a \in A$, and in turn produce the ranked list of APIs to be recommended for $p'$.

\subsection{Loss Function Formulation}

To solve the API recommendation task, we need to formulate the loss function that guides the training process of our ranking model.
We define a loss function $L(.)$ to evaluate the goodness of the compatibility score $f(p,a)$, and then find the optimal weight parameters that minimize $L(.)$. As mentioned, feature vectors $\mathbf{x}(p, a) = [x_1(p,a),\ldots,x_j(p,a)$, $\ldots,x_J(p,a)]$ are ranked according to $f(p,a)$. Thus, if the feature vectors with higher scores $f(p,a)$ are actually relevant (i.e., API $a$ is actually used by project $p$), the loss should be small; otherwise, the loss should be large.

In this work, we focus on a ranking loss function of the form $L(y(a >_p a'), f(a >_p a'))$, where $f(a >_p a')$ quantifies how likely API $a$ is more relevant to project $p$ than API $a'$, and $y(a >_p a')$ indicates whether $a$ is actually more relevant to $p$ than $a'$ (i.e., $y(a >_p a') = 1$ if $a >_p a'$, and $y(a >_p a') = -1$ otherwise). Accordingly, we can define the \emph{expected loss} $E$ over all possible project-API combinations as:
\begin{align}
E = \frac{1}{|P||A|^2} \sum_{p \in P} \sum_{a \in A} \sum_{a' \in A} L(y(a >_p a'), f(a >_p a')) \nonumber
\end{align}

By noticing that the training data $D$ defined in (\ref{eqn:pair_data}) contains only the API pairs $(a,a')$ such that $y(a >_p a') = 1$, and owing to the totality and anti-symmetry properties of a sound ranking, we can simplify the above formula as:
\begin{align}
E = \frac{1}{|D|} \sum_{(p,a,a') \in D} L(1, f(a >_p a'))
\end{align}

The above formulation by itself does not warrant a personalized total ordering. To achieve this, all three properties (i.e., totality, anti-symmetry, and transitivity) must be fulfilled. To this end, we can define $f(a >_p a')$ as:
\begin{align}
f(a >_p a') = f(p,a) - f(p,a')
\end{align}
which leads to the following loss:
\begin{align}
E = \frac{1}{|D|} \sum_{(p,a,a') \in D} L(1, f(p,a) - f(p,a'))
\end{align}

What then is a suitable choice for the loss function $L(.)$? In this work, we choose to use the \emph{squared hinge loss} $L(y, x) = \max(0, y(1-x))^2$, yielding the following expected loss:
\begin{align}
\label{eqn:hinge}
E = &\frac{1}{|D|} \sum_{(p,a,a') \in D} \max \left(0, 1 - (f(p,a) - f(p,a')) \right)^2
\end{align}
Intuitively, the above loss means that no penalty will be given to correct orderings (i.e., $f(p,a) > f(p,a')$), and a \emph{quadratic penalty} to incorrect orderings (i.e., $f(p,a) < f(p,a')$), depending on how far $f(p,a)$ is apart from $f(p,a')$. 

Quadratic penalty means that an incorrect ordering of APIs will get penalized higher (as compared to linear penalty). In other words, we are more stringent with incorrect ranking, which in principle would lead to a more robust model. Computationally, another merit of quadratic penalty is that we can compute the second derivative (also called curvature) of the loss function. As such, we can use second-order optimization methods (such as the Newton algorithm \cite{Lee2014}) to train the model faster. We further explain this in Section \ref{sec:training}.

To mitigate overfitting to the training data, we also add an L2 regularization term to the loss $E$, which leads to the regularized expected loss $R$:
\begin{align}
\label{eqn:hinge_L2}
R = &E + \frac{\lambda}{2} \sum_{j=1}^J \theta_j^2 
\end{align}
where  $\lambda > 0$ is the (user-defined) regularization parameter. Intuitively, adding the L2 regularization term serves to penalize large values of weight parameters $\theta_j$, which will have the effect of simplifying the ranking model and thus reducing the likelihood of overfitting. As such, performing the minimization of $E$ with the regularization term will provide us the simplest model that can fit the training data well.

It is also worth mentioning that the formulation of (\ref{eqn:hinge_L2}) can be viewed as a variant of the \emph{ranking support vector machine} (RankSVM) \cite{Joachims2002}. The conventional RankSVM, however, uses a linear hinge loss, which gives a less stringent linear penalty to incorrect orderings. Taking the analogy to classification task, it has been previously studied \cite{Lee2013} that using the squared hinge loss in SVM would yield better accuracy when $\lambda$ is large. In this case, underfitting would be less severe for the squared hinge loss, as it gives higher penalty than the hinge loss. The same argument applies to the ranking task, since RankSVM is ultimately equal to performing a binary classification on the pairwise feature differences $\Delta x_j = x_j(p,a) - x_j(p,a')$ \cite{Joachims2002}.

Finally, we note that the regularized loss $R$ is sound from the optimization viewpoint, as $R$ is a \emph{strictly convex} function. This means that there is a unique optimal solution for $\theta_j$, i.e., any local optimum found for $\theta_j$ will be the global optimum. The reason is that the second derivative of $R$ is always positive, that is, the Hessian matrix is positive definite \cite{Abe2002}. Thus, any gradient-based training method can be applied to arrive at a unique global optima. This constitutes another benefit of our approach over the regularized (linear) hinge loss used by the conventional RankSVM, which is not strictly convex.

\subsection{Efficient Training}
\label{sec:training}

While the regularized loss $R$ is strictly convex, the presence of a large number of API pairs $(a,a')$ would impose a high computational overhead. In particular, a na\"{i}ve computation of $R$ (as well as its derivatives) would have the time complexity of $O(\tilde{n} |D|^2)$ per iteration, which is quadratic with respect to the number of training triples $(p,a,a')$ in $D$. Here $\tilde{n}$ refers to the average number of nonzero features (i.e., $x_j(p,a) \neq 0$) per training triple. To mitigate this, we adopt an efficient \emph{truncated Newton} method as described in \cite{Lee2014}. The key idea is to first rewrite the Hessian (i.e, second derivatives) of the loss function in terms of matrix-vector product, and then exploit a special structure in the Hessian matrix for which some elements can be computed efficiently via an order-statistic tree \cite{Adelson-Velsky1962}. With this, we can bring the complexity down to $O(\tilde{n} |D| + |D| \log k)$, where $k$ is the number of relevance levels ($k = 2$ in our case, as we deal with binary relevance, i.e., whether or not an API is used by a project). Full details can be found in \cite{Lee2014}, and are not included in this paper for brevity.

\subsection{Ranking vs. Classification}

Why should we use a ranking approach instead of classification to address the recommendation problem? Indeed, one can use a classification method (e.g., binary SVM classifier) to distinguish whether an API is relevant to a project or not. However, such approach poses two main issues. First, the classification approach is built upon the premise that APIs that are not used by a project constitutes the \emph{negative} instances (i.e., will not be used by a project). Such assumption is inappropriate for the API recommendation task. In contrast, our ranking method assumes that such cases can either imply negative, or \emph{unobserved} (i.e., not yet explored in a project), instances. In this case, the ranking approach models the \emph{preferability} of APIs, i.e., if an API has been used by a project (i.e., positive instance), we assume that the project \emph{prefers} this API over all other negative and/or unobserved APIs.

Second, from a computational standpoint, the classification approach would suffer from the highly skewed distribution of positive and negative instances. This is because only a handful of APIs are actually used by a project (i.e., very few positive instances). In contrast, the ranking approach focuses on the preferability of APIs which exhibits the reversal property (i.e., if $a >_p a'$, then $a' <_p a$). As mentioned, RankSVM is equivalent to (binary) classification on a transformed feature space $\Delta x_j = x_j(p,a) - x_j(p,a')$. This leads to a transformed dataset whereby the class distribution is (automatically) balanced, which is easier to deal with.

%% file: feature.tex
\section{Feature Engineering}
\label{sec:feature}
In this section, we define features $x_j(p,a)$ that we use to train our personalized ranking model. We explore two groups of features: project features and API features.

\subsection{Project Features}\label{sec:pa}

To derive the project features, we first find the top-$k$ projects whose profiles are the most similar to the new project profile. APIs used in these top-$k$ projects are then used to calculate the API scores given the new project. We describe these two steps in the following subsections.

\subsubsection{Finding Top-$k$ Projects}\label{sec:patopk}

In order to find the top-$k$ projects, we need to measure the similarities between many projects. For two project profiles $p_1$ and $p_2$, we measure either the similarity of their textual descriptions or the similarity of their keywords. The detailed steps are as follows:

\vspace{0.2cm}
\begin{enumerate}
\item [i.] {\em Similarity of Textual Descriptions.} To compute the similarity between two textual descriptions, 
as mentioned in Section~\ref{sec:WebAPIRec}, we first convert each textual description to a VSM feature vector and then compute the similarity using cosine similarity between the two resultant feature vectors. The cosine similarity score corresponding to $p_1$ and $p_2$ is denoted as $Sim^{Text}(p_1,p_2)$.

\item [ii.] {\em Similarity of Keywords.} To compute the similarity between the keywords of $p_1$ and $p_2$, which we denote as $Sim^{Key}(p_1,p_2)$, we use the following formula:
\begin{align}\label{eqn:simkeypp}
Sim^{Key}(p_1,p_2) = \frac{|p_1^{Key} \cap p_2^{Key}|}{\sqrt{|p_1^{Key}| \times |p_2^{Key}|}}
\end{align}

where $p_1^{Key}$ and $p_2^{Key}$ corresponds to the set of keywords of $p_1$ and $p_2$ respectively. Also, $|p^{Key}|$ denotes the number of elements in the set $p^{Key}$. The numerator of the equation corresponds to the number of keywords that $p_1$ and $p_2$ have in common, while the denominator of the equation normalizes the similarity so that its score ranges from zero to one.\vspace{0.2cm}
\end{enumerate}

Notice that we separate descriptions and keywords so that we can distinguish their importance. It may be the case that the similarity of keywords is more important than the similarity of descriptions (and vice versa).


\subsubsection{Assigning Scores to APIs}

After a list of the top-$k$ projects is obtained (based on the similarity of textual descriptions or keywords), we analyze the set of APIs used in these projects. If an API is used by many of these top-$k$ projects, the API is likely more suitable for the new project. Considering a new project description $p'$ and project similarity measured in terms of textual descriptions, we assign a textual description based score to an API $a$ as:
\begin{align}
\mathit{CF}^{Text}(p',a,k) =
\frac{|\{p|p\in \mathcal{N}_k(p') \wedge y(p,a)=1\}|}{k}\label{eq:rescorehistory}
\end{align}
where $\mathcal{N}_k(p')$ denotes the top-$k$ projects of $p'$, and $y(p,a)$ indicates whether API $a$ is used by project $p$. The score $\mathit{CF}(p',a,k)$ ranges from 0 to 1. The higher the score is, the more likely API $a$ is suitable for the new project description $p'$. Similarly, we can measure project similarity in terms of keywords and compute $\mathit{CF}^{Key}(p',a,k)$.

We define our project features in terms of $\mathit{CF}(p',a,k)$. We consider different numbers of nearest neighbors $k$ and similarity definitions (i.e., description or keyword). We list these features in Table~\ref{tab:features}. The intuition behind this set of project features comes from the collaborative filtering concept, i.e., we are likely to find suitable APIs for a project by looking at other projects that are similar to it. The idea is that similar projects are likely to share common APIs because they share similar functionalities. Compare the descriptions of web application projects in Tables~\ref{tab:mashupEx} and~\ref{tab:mashupSim}. Both project descriptions contain words such as ``music'' and ``world'' and have a common keyword, i.e., ``music''. Note that the two projects share a common API namely ``Last.fm''.

\begin{table}[!t]
\scriptsize
  \centering
  \caption{Feature Definition}\label{tab:features}
  \vspace{-0.2cm}
  \begin{tabular}{lcl}
  \toprule
  {\bf Category}  & {\bf Feature} & {\bf Definition} \\
  \midrule
   \multirow{10}{*}{Project} & $x_1$ & $\mathit{CF}^{Text}(p',a,k)$ with $k=5$.\\
     & $x_2$ & $\mathit{CF}^{Text}(p',a,k)$ with $k=10$.\\
     & $x_3$ & $\mathit{CF}^{Text}(p',a,k)$ with $k=15$.\\
      &$x_4$ & $\mathit{CF}^{Text}(p',a,k)$ with $k=20$.\\
     & $x_5$ & $\mathit{CF}^{Text}(p',a,k)$ with $k=25$.\\
    \cline{2-3}
      &$x_6$ & $\mathit{CF}^{Key}(p',a,k)$ with $k=5$.\\
      &$x_7$ & $\mathit{CF}^{Key}(p',a,k)$ with $k=10$.\\
      &$x_8$ & $\mathit{CF}^{Key}(p',a,k)$ with $k=15$.\\
      &$x_9$ & $\mathit{CF}^{Key}(p',a,k)$ with $k=20$.\\
     &$x_{10}$ & $\mathit{CF}^{Key}(p',a,k)$ with $k=25$.\\
  \midrule
   \multirow{2}{*}{API} & $x_{11}$ & $Sim^{Text}$($p'$,$a$)\\
   \cline{2-3}
    &  $x_{12}$ & $Sim^{Key}$($p'$,$a$)\\
  \bottomrule
  \end{tabular}
  \vspace{-0.2cm}
\end{table}

\begin{table}[!t]
  \scriptsize
  \centering
  \caption{Another Sample Web Application Project Profile}\label{tab:mashupSim}
  \vspace{-0.2cm}
  \begin{tabular}{lp{1.9in}}
  \toprule
  \multicolumn{2}{c}{\bf Sound Shelter - An electronic music discovery engine} \\
  \midrule
   {\bf Long Description} &  Sound Shelter is an electronic music discovery engine. We listen to the opinions of the top taste makers from across the planet to bring you the world's best House, Techno, Disco, Dubstep and Soul Jazz releases. Our powerful web technology collects recommendations from music lovers all over the world to give you the best new releases across all electronic and soul jazz.\\
   {\bf Keywords} & music, search, recommendations\\
   {\bf APIs} & Last.fm, Discogs, Echo Nest, Spotify Metadata, Juno Download\\
  \bottomrule
  \end{tabular}
  \vspace{-0.5cm}
\end{table}

\subsection{API Features}\label{sec:aa}

We compare profiles of different APIs with a new project profile. For each API, we compute scores corresponding to the similarity between the API profile and the new project profile. For an API $a$ and a new project $p'$, we either measure the similarity of their textual descriptions or the similarity of their keywords. We consider these two similarity measures as our API features and list them in Table~\ref{tab:features}. The detailed steps to compute the similarity measures are as follows:

\begin{enumerate}
\item [i.] {\em Similarity of Textual Descriptions.} To compute a similarity score between an API's and a new project's textual descriptions, we convert these textual descriptions into vectors of weights following similar steps when computing similarity of textual descriptions between two projects in Section~\ref{sec:patopk}. We then compute the cosine similarity between the API and the new project feature vectors. 
We denote the cosine similarity between an API $a$ and a new project $p'$ as $Sim^{Text}$($p'$,$a$).\vspace{0.2cm}

\item [ii.] {\em Similarity of Keywords.} To compute a similarity score between the set of keywords for API $a$ and the set of keywords for the new project description $p'$, we follow Equation~\ref{eqn:simkeypp}. We denote the keywords similarity of an API $a$ and a new project $p'$ as $Sim^{Key}$($p'$,$a$).

\end{enumerate}

The rationale behind using similarity between a project and an API as features is that a project profile should explain the project functionality while an API profile should explain the API functionality. Thus, an API that is more similar to a project is likely to be more suitable for the project since they are likely to share similar functionality. Consider the project profile in Table~\ref{tab:mashupEx} and the API profile in Table~\ref{tab:apiEx}, both the project and API descriptions contain words such as ``fm'', ``music'' and ``artists'', and share a common keyword, i.e., ``music''. In this case, we can say that the API is likely to be  usable for the project.

%% file: experiment.tex
\section{Experiments}
\label{sec:experiment}


\subsection{Dataset, Metrics, and Settings}\label{sec:dataset}

\noindent{\bf Dataset.} ProgrammableWeb's site contains the profiles of more than 17,000 APIs and more than 7,000 web application projects. {However, ProgrammableWeb specifies that a number of APIs and projects are no longer offered by the providers. ProgrammableWeb explicitly labels such APIs and projects as deprecated. This allow us to delete corresponding APIs and projects automatically. We delete these phased out APIs and projects and focus on those that are available for use.} After we delete these phased out APIs, we are left with 9,883 APIs and 4,315 projects which we use for this study. The goal of our experiment is to investigate whether {\tt {WebAPIRec}} can return correct APIs given the profile of a project. The ground truth APIs of a project are the APIs that are specified in the project's page on the ProgrammableWeb's site. Note that these APIs are used by the project and thus prove to be useful APIs. 

{In our preliminary investigation on some projects in ProgrammableWeb, we notice that their textual descriptions sometimes explicitly mention the API names that are used by the projects. We remove all mentions of these API names from the project description. This is necessary to ensure that the description contains no mention about the correct APIs. The removal process is fully automatic since the mentions of API names in project textual descriptions are exact and thus removing them simply requires us to perform a simple textual search and replace procedure.}

{One may ask whether the ground truth obtained from ProgrammableWeb is reliable. Due to the large size of the dataset, it is impossible for us to know whether all the ground truth is valid. To mitigate this threat to the validity of our findings, one of the authors have manually checked the correctness of the ground truth for a random subset of 353 projects to achieve statistically significant result at a confidence level of 95\% and margin of error of 5. We found that the ground truth is correct. We consider a ground truth to be correct if used APIs’ functionalities do not conflict with a project’s description. Conflict happens when we cannot find reasons on why an API would be used by a project given its description. On the random subset, we find that no such conflict occurs.}

\vspace{0.2cm}\noindent{\bf Evaluation Metrics.} To evaluate our approach, we consider several popular evaluation metrics Hit@N, Mean Average Precision (MAP), MAP@N, and Mean Reciprocal Rank (MRR). These metrics have been used before in many previous studies~\cite{shaowei/icpc2014,RaoK11,ZhouZL12,SahaLKP13,SunLWJK10,RAN07,ThungWLL13,manning2008introduction}. We elaborate these metrics below:

\begin{itemize}

\item{\textbf{Hit@$N$:}} This metric counts the percentage of ranked lists produced when recommending APIs to projects, where at least one correct API exists at the top $N$ results. In this work, we use $N=5$ and $10$.

\item{\textbf{Mean Average Precision (MAP):}} MAP is a popularly used IR metric to evaluate the ranking results. It exhibits a \emph{top-heaviness} trait,  putting higher penalties for incorrect ordering at the top ranked APIs \cite{manning2008introduction}. To compute MAP, for each ranked list returned for a project, we first compute the average precision (AP):
\begin{align}
\label{eqn:avg_prec}
 AP = \frac{ \sum\limits_{i=1}^{M} P(i)\times rel(i)}{\sum\limits_{i=1}^{M} rel(i)}
\end{align}
where $M$ is the number of retrieved APIs, $rel(i)$ is a binary value that represents whether the $i^{th}$ retrieved API is correct or not, and $P(i)$ is the precision at position $i$ of the ranked list. $P(i)$ is defined as:
\begin{align}
 P(i) = \frac{\#\text{Correct APIs at top i positions}}{\text{i}}
\end{align}
In turn, MAP is the mean of the APs over all projects.

\item{\textbf{MAP@$N$:}} This is the same as MAP, except that we replace $M$ in equation~(\ref{eqn:avg_prec}) to $N$, where $ N \ll M$. We use this metric to account for limited attention bandwidth, i.e., a developer can look only at a limited number ($N$) of APIs. In this work, we use $N=5$ and $10$.

\item{\textbf{Mean Reciprocal Rank (MRR):}} The reciprocal rank of a ranked list is the inverse of the rank of the first correct API in the ranked list. The mean reciprocal rank takes the average of the reciprocal ranks of all ranked lists produced when recommending APIs to projects. For a set of projects P, MRR is defined as:
\begin{align}
MRR=\frac{1}{|P|}\sum_{i=1}^P{\frac{1}{rank_i}}
\end{align}
where $rank_i$ is the rank of the first correct API.


\end{itemize}

\noindent{\bf Experiment Setting.} We use 10-fold cross validation to evaluate our approach. That is, we first divide the projects into 10 mutually exclusive parts (i.e., folds), We then use 9 parts to train the weight parameters of our personalized ranking model (i.e., training set), and use the remaining part to evaluate the performance of our model (i.e., testing set). We repeat the process 10 times using 10 mutually exclusive testing sets. We aggregate the performance across the 10 folds and report the average scores. All experiments were conducted on an Intel(R) Xeon CPU E5-2667 @2.90 GHz PC with Linux CentOS operating system. For all experiments, we set the regularization parameter $\lambda$ of our ranking method to $1$.

\subsection{Baseline Methods}\label{sec:baselines}
We use the following baselines to gauge our {\tt {WebAPIRec}} approach:
\begin{enumerate}
\item {\em ProgrammableWeb Search Functionality.} For this baseline, we type the query in ProgrammableWeb search box and check whether the recommended APIs match the APIs that were actually used by the project. We consider three variants of this baseline approach: the first variant only uses the project description ({\tt PW$^{Text}$}), only uses the project keywords (i.e., tags) ({\tt PW$^{Key}$}), and both ({\tt PW$^{Text+Key}$}). {Note that we do not perform any preprocessing for ProgrammableWeb input since developers would also not do so. Moreover, ProgrammableWeb might perform it internally. }

\item {\bf \tt Exemplar$^{API}$.} {This is an adapted version of McMillan \emph{et al.}'s work~\cite{mcmillan2012exemplar}. They proposed Exemplar, a search engine for relevant applications. In our work, we treat an API as an application and search ``relevant applications'' using project profile. To use Exemplar in our setting, we need to remove its source code analysis component, since our scenario only involves text as input. We note that many APIs, including the web APIs considered in this work, do not come with source code. After this treatment, Exemplar approach is equivalent to an approach that computes VSM text similarity between project and API descriptions, and uses the resultant similarity scores to rank APIs. Since Exemplar code is not made publicly available, we reimplemented it based on the authors’ description in the paper.}


\item {\bf \tt PopRec}. This is a popularity-based recommendation  baseline. We define popularity of an API as the number of times the API has been used on the list of projects in the training data. Therefore, a more popular API will have a higher rank in the recommendation list output by {\tt PopRec}. In this approach, the same list of APIs will be recommended to each project in the evaluation data. In other words, the recommendation is not personalized. The top-50 popular APIs are shown in Table~\ref{tab:top50API}.
\end{enumerate}

\begin{table}[!t]
	\scriptsize
	\centering
	\caption{ Top-50 APIs in ProgrammableWeb}\label{tab:top50API}
	\begin{tabular}{p{8cm}}
		\toprule
		{\bf API Names}\\
		\midrule
		Google Maps, Twitter, YouTube, Twilio, Facebook, Amazon Product Advertising, Twilio SMS, eBay, Last.fm, Microsoft Bing Maps, DocuSign Enterprise, Google App Engine, foursquare, Google Homepage, Box, GeoNames, del.icio.us, Amazon S3, Shopping.com, Amazon EC2, Concur, indeed, Instagram, Google AdSense, LinkedIn, Salesforce.com, Freebase, Facebook Graph, Yelp, Spotify Metadata, Wikipedia, Google Earth, Bing, Bit.ly, Yahoo BOSS, Google AJAX Libraries, Google Analytics, Google Geocoding, Lyricsfly, Google Ajax Feeds, Google Translate, MusicBrainz, Panoramio, Bing Maps, Oodle, SoundCloud, PayPal, Zillow, Google Calendar, Facebook Social Plugins \\
		\bottomrule
	\end{tabular}
	\vspace{-0.4cm}
\end{table}

{For all baselines, we simulate how developers search APIs as observed from the ProgrammableWeb interface. This makes our baselines meaningful since it reflects real world scenario.  For all approach (including ours), if two APIs have the exact ranking score, we randomly break the tie.} 

\subsection{Key Results and Analysis}

{\bf{\em RQ1:} How Effective is Our Approach in Recommending APIs to Projects?}
We evaluate the extent our approach {\tt {WebAPIRec}} is effective to recommend APIs to projects. We compare our approach with the baselines in Section~\ref{sec:baselines}. Evaluation is done via a 10-fold cross validation procedure, and for each project, we use {\tt {WebAPIRec}} and the baselines to recommend APIs based on the project profile.

\input{rq1}

{\bf{\em RQ2:} What is the Contribution of Each Feature in Our Ranking Model?}
We evaluate the contribution of each feature in our approach. The goal is to know which features are more important. To this end, we use the weight parameters $\theta_j$ in our model. Features with higher weight values are considered to have higher contributions and are thus more important. As we perform 10-fold cross validation, we average the feature weights across 10 folds. This gives us the average contribution of each feature. We then report these average weights to indicate which features are the most important.

\input{rq2}

{\bf{\em RQ3:} What is the Impact of Training Size to the Effectiveness of Our Approach?}
We investigate the effect of training size to the effectiveness of our approach. To this end, we keep the same 10\% of data as our testing set, but use different percentages of data as training set: 10\%, 20\%, 30\%, $\ldots$, 80\%. By keeping the same set of evaluation data, we ensure that the impact of training size is comparable. For each percentage of training data, we report the average performance. 
	
\input{rq3}

{\bf{\em RQ4:} How Efficient is Our Approach During Its Training and Deployment Phases?}
The efficiency of {\tt {WebAPIRec}} affects its practical use. Thus, we investigate the time it takes for {\tt {WebAPIRec}} to learn its weights from training data and the time it takes to recommend APIs to a project. Firstly, to measure training efficiency, we log the training time for each CV fold and report the averaged (training) time over 10 folds. Secondly, we measure recommendation efficiency by computing the total time required to predict on the 10 testing sets, and dividing it with 10 times the total number of projects.

\input{rq4}

\subsection{Threats to Validity}
\input{threats} 

%% file: rq1.tex
Table~\ref{tab:effectiveness} illustrates the effectiveness of our approach in comparison with the baselines. Our approach achieves Hit@5, Hit@10, MAP@5, MAP@10, MAP, and MRR scores of 0.840, 0.880, 0.697, 0.687, 0.626, and 0.750, respectively. Both MAP@5 and MAP@10 scores are lower than Hit@5 and Hit@10 scores. This indicates that in the top-N, for most cases, not all APIs are relevant, but at least one of them are. Based on Hit@5 results, we find that for 84.0\% of the projects, a correct API used to implement a project is among the top-5 APIs returned by {\tt WebAPIRec}. Clearly, {\tt WebAPIRec} outperforms the baselines that use ProgrammableWeb native search functionality. Measured either by Hit@5, Hit@10, MAP or MRR, {\tt WebAPIRec} performs better than {\tt PW$^{Key}$}, which is the best performing baseline from ProgrammableWeb. {{\tt PW$^{Text+Key}$}, which has the largest number of words among the ProgrammableWeb baselines, performs the worst. In fact, we observe a consistent reduction in performance as number of words increases. We hypothesize that ProgrammableWeb uses boolean {\em and} operation in its search engine, thereby returning only APIs whose profiles contain all words in the query. Our manual investigation suggests it is likely the case.}

{\tt WebAPIRec} also outperforms {\tt Exemplar$^{API}$} and {\tt PopRec}. 
The strongest baseline is {\tt PopRec}, which achieves Hit@5, Hit@10, MAP@5, MAP@10, MAP, and MRR scores of 0.591, 0.675, 0.414, 0.417, and 0.476, respectively. {\tt WebAPIRec} clearly improves significantly upon this baseline by 42.1\% in terms of Hit@5.

We show an example recommendation from our approach in Table~\ref{tab:exampleRec}. Here, we recommend APIs for Yamusica. Three of our recommendations are correct: Bandsintown, Last.fm, and Google Maps. The baselines can only recommend less than 3 correct APIs: ProgrammableWeb does not return any APIs; {\tt PopRec} can identify Google Maps because it is in the set of top-10 APIs as shown in Table~\ref{tab:top50API}. {\tt Exemplar$^{API}$} can correctly recommends one API, i.e. Bandsintown, since its description and keywords are the most similar with Yamusica. {\tt WebAPIRec} recommends the three APIs largely because they are used by similar projects.

\begin{table}[!t]
	\scriptsize
	\centering
	\caption{Example Recommendation}\label{tab:exampleRec}
	\vspace{-0.2cm}
	\begin{tabular}{lp{1.9in}}
		\toprule
		\multicolumn{2}{c}{\bf Yamusica} \\
		\midrule
		{\bf Long Description} &  Yamusica brings you the latest information on live music events, concerts and venues around the world. Find out when your favorite artist is coming to a town near you! \\
		{\bf Keywords} & Mapping, Events, Music, Tickets\\
		{\bf APIs} & Google Maps, Last.fm, GeoNames, Spotify Echo Nest, Bandsintown\\
		{\bf Recommendations} & Bandsintown, Last.fm, Eventbrite, SeatGeek, Google Maps, IP Location, StubHub, Facebook, Active.com, Seatwave, OpenWeatherMap\\
		\bottomrule
	\end{tabular}
\end{table}



%

\begin{table}[!t]
    \fontsize{6.5}{6}\selectfont
	\centering
	\caption{Effectiveness of Our Approach}\label{tab:effectiveness}
	\begin{tabular}{lcccccc}
	    \toprule
		{\bf Approach} & {\bf Hit@5} & {\bf Hit@10} & {\bf MAP@5} & {\bf MAP@10}  & {\bf MAP} & {\bf MRR} \\
		\midrule
		{\tt PW$^{Text}$} & 0.005 & 0.005 & 0.004 & 0.004 & 0.004 & 0.004\\
		{\tt PW$^{Key}$} & 0.046 & 0.047 & 0.041 & 0.042 & 0.042 & 0.038\\
		{\tt PW$^{Text+Key}$} & 0.001 & 0.001 & 0.001 & 0.001 & 0.001 & 0.001\\
		{\tt Exemplar$^{API}$} & 0.184 & 0.236 & 0.113 & 0.114 & 0.096 & 0.147\\
		{\tt PopRec} &  0.591 & 0.675 & 0.414 & 0.417 & 0.363 & 0.476\\
		{\tt WebAPIRec} & 0.840 & 0.880 & 0.697 & 0.687 & 0.626 & 0.750\\
		\bottomrule
	\end{tabular}
\end{table} 

%% file: rq2.tex
Table~\ref{tab:fcontrib} shows the contribution of each feature to the effectiveness {\tt WebAPIRec}. The most important feature based on our model weights is feature $x_3$, which is our project feature that considers top-15 most similar projects in terms of their descriptions. It suggests that 15 nearest neighbors is the optimal number of neighbors. Using too few neighbors may hurt recommendation accuracy as pertinent information from other useful neighbors may be missed. On the other hand, using too many neighbors may increase noise since irrelevant neighbors may be included. Our model gives less weights for features extracted from too few neighbors (e.g., $x_2$) or too many neighbors (e.g., $x_4$). Similar observation is reflected from the weights of features $x_6$-$x_{10}$, which measure keyword similarities. We observe that the highest weight is also given to the feature representing top-15 most similar projects (i.e., $x_8$) and less weights are given to features that are extracted from either too few neighbors (e.g., $x_7$) or too many neighbors (e.g., $x_9$). Meanwhile, we observe that the API features give moderate contribution to the performance of {\tt WebAPIRec}.


\begin{table}[!t]
    \scriptsize
	\centering
	\caption{Contributions of Individual Features}
	\label{tab:fcontrib}
	\begin{tabular}{ccc}
	    \toprule
		{\bf Feature} & {\bf Definition} & {\bf Weight} \\
		\midrule
		$x_1$ & $\mathit{CF}^{Text}(p',a,5)$ & 0.397  \\
		$x_2$ & $\mathit{CF}^{Text}(p',a,10)$ & 0.329 \\
		$x_3$ & $\mathit{CF}^{Text}(p',a,15)$ & 2.403 \\
		$x_4$ & $\mathit{CF}^{Text}(p',a,20)$ & 1.135 \\
		$x_5$ & $\mathit{CF}^{Text}(p',a,25)$ & 0.785 \\
		\midrule
		$x_6$ & $\mathit{CF}^{Key}(p',a,5)$ &-0.058 \\
		$x_7$ & $\mathit{CF}^{Key}(p',a,10)$ & 0.425\\
		$x_8$ & $\mathit{CF}^{Key}(p',a,15)$ & 0.963\\
		$x_9$ & $\mathit{CF}^{Key}(p',a,20)$ & 0.677\\
		$x_{10}$ & $\mathit{CF}^{Key}(p',a,25)$ & 0.459 \\
		\midrule
		$x_{11}$ & $Sim^{Text}$($p'$,$a$) & 0.600 \\
		$x_{12}$ & $Sim^{Key}$($p'$,$a$) & 0.497 \\
		\bottomrule
	\end{tabular}
    \vspace{-0.3cm}
\end{table}

%% file: rq3.tex
Table~\ref{tab:varySize} shows the effectiveness of {\tt WebAPIRec} when we vary the training size. We notice that the performance of our approach increases as the size of the training data increases. Moreover, the improvement direction is always consistent among different evaluation measures, meaning that the performance never drops as we increase the training size. Moreover, even when the size of the training data is only 10\%, we can still successfully recommend correct APIs in the top-5 positions for 75.6\% of the projects.


\begin{table}[!t]	
	\fontsize{7}{8}\selectfont
	\centering
	\caption{Varying Training Size}\label{tab:varySize}
	\begin{tabular}{ccccccc}
		\toprule
		{\bf Size} & {\bf Hit@5} & {\bf Hit@10} & {\bf MAP@5} & {\bf MAP@10} & {\bf MAP} & {\bf MRR}\\
		\midrule
		10\% & 0.756 & 0.818 & 0.618 & 0.610 & 0.540 & 0.673\\
		20\% & 0.782 & 0.821 & 0.646 & 0.638 & 0.574 & 0.696\\
		30\% & 0.802 & 0.842 & 0.662 & 0.653 & 0.588 & 0.713\\
		40\% & 0.820 & 0.855 & 0.672 & 0.662 & 0.600 & 0.725\\
		50\% & 0.831 & 0.865 & 0.680 & 0.671 & 0.608 & 0.734\\
		60\% & 0.834 & 0.872 & 0.685 & 0.676 & 0.614 & 0.739\\
		70\% & 0.835 & 0.876 & 0.688 & 0.679 & 0.617 & 0.741\\
		80\% & 0.836 & 0.879 & 0.694 & 0.684 & 0.623 & 0.746\\
		90\% & 0.840 & 0.880 & 0.697 & 0.687 & 0.626 & 0.750\\
		\bottomrule
	\end{tabular}
	\vspace{-0.3cm}
\end{table}

%% file: rq4.tex
Table~\ref{tab:efficiency} shows the consolidated results. 
On average, {\tt WebAPIRec} only needs about three minutes to train a model, and 0.0013 seconds to recommend a list of APIs to a project. In practice, training only needs to be performed once or occasionally (when the training data changes significantly). The results show that {\tt WebAPIRec} is efficient.

\begin{table}[!t]
  \scriptsize
  \centering
  \caption{ Computational Time of {\tt WebAPIRec}}\label{tab:efficiency}
  \begin{tabular}{cc}
  \toprule
	{\bf Phase} & {\bf Average Time} \\
  \midrule
 	Training & 179.7 seconds \\
 	Deployment & 0.0013 seconds \\
  \bottomrule
  \end{tabular}
  \vspace{-0.4cm}
\end{table} 

%% file: threats.tex

\vspace{0.2cm}\noindent{\bf Threats to Internal Validity.} It relates to experimental errors and biases. We have double-checked the correctness of our codes. Still, there could be bugs that we  miss. Also, some APIs and projects in ProgrammableWeb are no longer in service. As mentioned in Section~\ref{sec:dataset}, we have cleaned our dataset by removing these APIs and projects. We have also removed explicit mentions of API names from project descriptions. {Another potential threat is related to project descriptions itself. The descriptions are likely written post-implementation and thus may not reflect pre-implementation descriptions. Unfortunately, there is no public dataset containing pre-implementation descriptions and used APIs. However, ProgrammableWeb descriptions are typically brief whereas requirement documents (i.e., examples of pre-implementation descriptions) are much more detailed and thus are expected to lead to better performance (i.e., due to richer information).}

\vspace{0.2cm}\noindent{\bf Threats to External Validity.} It relates to the generalizability of our results. We have evaluated our method on a dataset comprising 9,883 APIs and 4,315 projects. We believe these are sufficiently large numbers of APIs and projects. Still, all APIs and projects come from ProgrammableWeb. 
{In the future, we plan to mitigate the threats to external validity further by investigating additional APIs and projects. Note that our approach can potentially be used for non-web APIs, provided that the same set of information exists. We plan to explore how our approach works for non-web APIs in the future. 
} 

\vspace{0.2cm}\noindent{\bf Threats to Construct Validity.} It relates to the suitability of our evaluation metrics. In this work, we have used Hit@N, MAP (as well as MAP@$N$), and MRR, which have been well-established in IR community and many past software engineering studies~\cite{shaowei/icpc2014,RaoK11,ZhouZL12,SahaLKP13,SunLWJK10,RAN07,ThungWLL13}. Thus we believe there is little threat to construct validity. 

%% file: related.tex
\section{Related Work}
\label{sec:related_works}


\vspace{0.2cm}\noindent{\bf Studies on Method Recommendation.} Thummalapenta and Xie ~\cite{ThummalapentaX07} proposed an approach to recommend code snippet. Their approach queries a code search engine (i.e., Google Code) to return code examples. These examples are then used to infer a sequence of method invocations for converting an object from one type to another.
Robbes and Lanza~\cite{RobbesL10} proposed a technique that improves code auto-completion by using recorded program history. Hindle \emph{et al.}~\cite{HindleBSGD12} investigated the ``naturalness'' of software, and proposed a code auto-completion feature by building a statistical language model. Kawaguchi \emph{et al.}~\cite{KawaguchiYUFKNI99a} and Lee \emph{et al.}~\cite{LeeRHK10} developed tools that are able to detect code clone in real time. These tools can also potentially be used for code auto-completion.

Chan \emph{et al.}~\cite{ChanCL12} proposed an approach to recommend API methods given textual phrases. Their approach was extended by Thung \emph{et al.}~\cite{ThungWLL13}, who recommend API methods given a feature request. Chan \emph{et al.}'s approach requires precise textual queries, whereas Thung \emph{et al.}'s approach is more robust to noisy textual queries.
Robillard \emph{et al.}~\cite{Robillard05} developed Suade, which takes as input a set of program elements and outputs another set of program elements that would likely be interesting to the developers. 
Different from the above studies, in this work we do not recommend API methods; rather, we recommend the APIs. Our work is thus complementary with the above studies. Developers can first use our approach to infer the web APIs relevant to a project, and then adapt some of the tools in the above studies to recommend relevant methods from the APIs.


\vspace{0.2cm}\noindent{\bf Studies on API Recommendation.} Teyton \emph{et al.}~\cite{TeytonFB12} proposed an approach that creates a library migration graph by analyzing library migrations performed by a large number of projects. This graph can be used to help developers decide appropriate libraries to migrate to. Teyton \emph{et al.}'s work and our work have different yet complementary goals: recommending libraries to migrate old libraries of existing projects vs. recommending libraries to new projects. 
Thung \emph{et al.}~\cite{ThungLL13} devised an approach that takes as input APIs that a project uses and recommends additional relevant APIs. Different from the current work, this approach does not take as input the profile of a {\em new} project. Instead, it requires developers to input APIs that are used by an {\em existing} project. 
It does not employ any text mining solution, since no text data is involved. In contrast, {\tt {WebAPIRec}} employs text mining and does not require information about APIs that are or will be used in a project. {\tt {WebAPIRec}} can thus be used in the {\em initial development stage}, when only the requirement of a project is known. Also, our work complements the work in~\cite{ThungLL13}, as developers can pick suitable APIs from our recommendation and put these APIs as input to the method in~\cite{ThungLL13} to get additional recommendations.



\vspace{0.2cm}\noindent{\bf Studies on ProgrammableWeb Dataset.} There exist a number of studies on the ProgrammableWeb dataset~\cite{han2014mining,weiss2010modeling,yu2009innovation}. These studies tried to characterize the structure and evolution of various networks created from the APIs and projects that are listed on ProgrammableWeb. Various network properties such as power-law, long-tail, small-world, etc. were investigated in these studies. For example, a recent work by Han \emph{et al.}~\cite{han2014mining} analyzed whether or not networks created from APIs, projects and their tags in ProgrammableWeb have power-law properties. Different from the above studies, 
we are interested in recommending APIs on ProgrammableWeb.

\vspace{0.2cm}\noindent{\bf Studies on Text Mining for Recommending Developer Actions.} Almhana et al. propose to use multi objective optimization algorithm for bug localization~\cite{almhana2016recommending}. They define two optimization objectives for bug localization. The first one is maximizing both lexical and historical similarities and the second one is minimizing the number of recommended classes. Ye et al. have defined 6 similarity functions between bug reports and source codes that encode project domain knowledge~\cite{ye2014learning}. These similarities are input to their learning to rank approach. Given a new bug report, their approach ranks source code files in order of likelihood of them being the source of bug. Tian et al. use learning to rank approach to recommend developers for fixing issues described in bug reports~\cite{tian2016learning}.  Yang et al. combine word embedding and traditional information retrieval approach to recommend similar bug reports~\cite{yang2016combining}. Xia et al. combine bug report and developer based analysis to recommend developers that should be assigned to a bug report~\cite{xia2013accurate}.

\balance

%% file: conclusion.tex
\section{Conclusion and Future Work}
\label{sec:conclusion}

We have proposed {\tt {WebAPIRec}}, a recommendation system that takes as input a new project profile and recommends {web} APIs that are potentially relevant to the project. We have evaluated our approach on 9,883 {web} APIs and 4,315 projects in ProgrammableWeb.
{\tt {WebAPIRec}} achieves Hit@5, Hit@10, MAP@5, MAP@10, MAP, and MRR of 0.840, 0.880, 0.697, 0.687, 0.626, and 0.750, respectively. {\tt {WebAPIRec}} can thus successfully recommend correct {web} APIs in top-5 positions for 84.0\% of the projects. We have compared {\tt {WebAPIRec}} ProgrammableWeb's native search functionality, McMillan \emph{et al.}'s application search engine~\cite{mcmillan2012exemplar}, and popularity-based recommendation. {\tt {WebAPIRec}} always produce superior results.

As future work, we plan to analyze more APIs and more projects from additional data sources beyond ProgrammableWeb. {We also plan to consider context information to improve our approach (e.g., a word ``developer'' could mean either a real estate developer or a software developer). The context of a word can often be inferred from words appearing before or after the target word. To consider context information, we plan to employ deep learning (e.g., Word2Vec~\cite{mikolov2013distributed} ). Moreover, textual information from project profile in ProgrammableWeb may not contain all technical details and this may be a factor contributing to some inaccurate recommendations in our experiments. We plan to address this limitation by enriching descriptions of web APIs with information from other sources, e.g., online forums and Twitter feeds where users of web APIs share their experience and queries, and developers provide additional technical information to respond to user queries.} We also wish to extend our study to not only recommend APIs, but also suitable resources to help developers get started with the APIs. Last but not least, we wish to develop an approach that can provide rationales for recommended APIs (e.g., explaining why an API can be used for a given project).